\documentclass[twocolumn,showpacs,aps,amsmath,amssymb,floatfix]{revtex4}
\usepackage{epsf}
\begin{document}
\title{Random Geometric Series}
\author{E.~Ben-Naim}
\affiliation{Theoretical Division and Center for Nonlinear Studies,
Los Alamos National Laboratory, Los Alamos, New Mexico 87545, USA}
\author{P.~L.~Krapivsky}
\affiliation{Center for Polymer Studies and Department of Physics,
Boston University, Boston, Massachusetts 02215, USA}
\begin{abstract}
  Integer sequences where each element is determined by a previous
  randomly chosen element are investigated analytically. In
  particular, the random geometric series $x_n=2x_p$ with \hbox{$0\leq
  p\leq n-1$} is studied. At large $n$, the moments grow
  algebraically, $\langle x_n^s\rangle\sim n^{\beta(s)}$ with
  $\beta(s)=2^s-1$, while the typical behavior is $x_n\sim n^{\ln
  2}$. The probability distribution is obtained explicitly in terms of
  the Stirling numbers of the first kind and it approaches a log-normal
  distribution asymptotically.
\end{abstract}

\maketitle

\section{Introduction}

Integer sequences are ubiquitous in pure and applied mathematics,
physics, and computer science \cite{cg}. While traditional integer
sequences are deterministic, there is a growing interest in stochastic
counterparts of fundamental sequences and their relevance to
disordered or random systems.  For example, the random Fibonacci
sequence $x_n=x_{n-1}\pm x_{n-2}$ \cite{hf,dv,sk} has links with various
topics in condensed matter physics, dynamical systems, products of
random matrices, etc. (see e.g. \cite{bl,cpv,jml,dh,ob,hoz}).

Random integer sequences are conceptually simple, yet they exhibit a
complex phenomenology resulting from the memory generated by the
stochastic recursion law. For the random Fibonacci sequence, the
typical behavior is $x_n\sim e^{\lambda n}$ with the intriguing
Lyapunov exponent $\lambda=0.12397559$; furthermore, the distribution
of the ratio $x_n/x_{n-1}$ has singularities at every rational value
\cite{dv,sk} and the model exhibits a remarkably intricate spectrum
\cite{hoz}. Also, the typical growth is different than the average
growth as characterized by the moments $\langle x_n^s\rangle$.

Another form of randomness in which an element in the series depends
on two previous elements, at least one of which is chosen randomly,
was introduce recently \cite{bk}. For example, the stochastic
Fibonacci-like series defined recursively by the rule
$x_n=x_{n-1}+x_p$ with $p$ randomly chosen between $0$ and $n-1$,
exhibits the typical growth $x_n\sim \exp(\lambda\sqrt{n})$ with the
(numerically calculated) Lyapunov exponent $\lambda=1.889$. The
moments exhibit multiscaling \cite{bk,krt} and the probability
distribution becomes log-normal asymptotically.  Small variations in
the recurrence rule can lead to substantial changes in the sequence
characteristics --- the growth law may be algebraic, log-normal, or
exponential and the distribution may or may not exhibit multiscaling
of the moments.  In this article, we show that much of this
phenomenology is captured by even simpler stochastic series, for which
a more detailed analytical treatment is feasible.

\section{Random Multiplicative Series}

We consider recursively defined series where an element is determined
by a {\em single} previously chosen element. A natural starting point
is the random geometric series
\begin{equation}
\label{rule-mult}
x_n=2x_p,
\end{equation}
when $n\geq 1$ and $x_0=1$.  The index $p$ is chosen randomly between
$0$ and $n-1$ at each step. The first element is $x_0=1$. For example,
for $n\leq 2$, there are two, equally probable sequences:
${x_n}={1,2,2}$ or ${1,2,4}$. The series may not necessarily be
monotonic.

\subsection{The Moments}

The moments $\langle x_n^s\rangle$ can be obtained analytically. They
obey the recursion relation
\begin{equation}
\label{mom-rec}
\langle x_n^s\rangle=\frac{2^s}{n}
\sum_{p=0}^{n-1}\langle x_p^s\rangle
\end{equation}
for $n\geq 1$, with $\langle x_0^s\rangle=1$. This recursion is
solved using the generating function
$M(s,z)=\sum_{n=0}^\infty\langle x_n^s\rangle\,z^{n}$. The
recursion relation (\ref{mom-rec}) leads to the ordinary
differential equation $\frac{dM}{dz}=\frac{2^s}{1-z}M$ subject to
the boundary condition $M(s,0)=1$. Expanding the solution
\begin{equation}
M(s,z)=(1-z)^{-2^s}
\end{equation}
in powers of $z$ gives the moments
\begin{equation}
\label{mom-sol}
\langle x_n^s\rangle=\frac{\Gamma(n+2^s)}{\Gamma(2^s)\,\Gamma(n+1)}.
\end{equation}

The first and the second moment are given by simple polynomials
\begin{eqnarray}
\langle x_{n}\rangle =n+1,\quad
\langle x_{n}^2\rangle=\frac{(n+1)(n+2)(n+3)}{6}.
\end{eqnarray}
Generally, there is a series of special values of $s$ for which the
moments are polynomial in $n$. For $2^s=k$ with $k$ being an integer,
the moments are
\begin{eqnarray}
\langle x_n^{\ln k/\ln 2}\rangle=\frac{(n+1)(n+2)\cdots(n+k-1)}{(k-1)!}.
\end{eqnarray}
Asymptotically, all moments grow algebraically
\begin{equation}
\label{moments}
\langle x_n^s\rangle\simeq A(s)\,n^{\beta(s)}
\end{equation}
with $\beta(s)=2^s-1$ and $A(s)=1/\Gamma(2^s)$. Therefore, the moments
exhibit a multiscaling behavior characterized by the nonlinear
spectrum of exponents $\beta(s)$.  

\subsection{The Probability Distribution}

The probability distribution of the random variable $x_n$ and its
typical behavior are obtained by considering a closely related random
sequence.  Since the spectrum of possible values for $x_n$ is $2^m$
with integer $m\geq 0$, we study the variable $m_n=\log_2 x_n$. The
corresponding random additive series obeys the recursion
relation
\begin{equation}
\label{rule-add}
m_n=m_p+1
\end{equation}
with $m_0=0$ and a randomly chosen $0\leq p\leq n-1$. Generally,
$1\leq m_n\leq n$ for $n\geq 1$. The probability distribution
$P_{n,m}={\rm Prob}(x_n=2^m)$ satisfies
\begin{equation}
\label{Pnm-recur}
P_{n,m}=\frac{1}{n}\sum_{l=0}^{n-1} P_{l,m-1}
\end{equation}
for $n\geq 1$ and $P_{0,m}=\delta_{m,0}$. From this recursion, one readily
obtains $nP_{n,m}-(n-1)P_{n-1,m}=P_{n-1,m-1}$, thereby recasting
(\ref{Pnm-recur}) into
\begin{equation}
\label{dis-rec}
P_{n,m}=\frac{n-1}{n}P_{n-1,m}+\frac{1}{n}P_{n-1,m-1}\,.
\end{equation}
To tackle this recursion it is convenient to eliminate the
denominator.  The modified distribution $G_{n,m}=n!P_{n,m}$ satisfies
the recursion
\begin{equation}
\label{mod-rec}
G_{n,m}=(n-1)G_{n-1,m}+G_{n-1,m-1}
\end{equation}
with $G_{0,m}=\delta_{m,0}$. The very same recurrence generates ${n\brack
  m}$, the Stirling numbers of the first kind \cite{gkp}. These numbers are
closely related to the binomial coefficients and appear in numerous
applications \cite{kr,bkr,lg}.

Thus $G_{n,m}={n\brack m}$, and the probability distribution is expressed in
terms of these special numbers as follows:
\begin{equation}
\label{dis-sol}
P_{n,m}=\frac{1}{n!}\,{n\brack m}.
\end{equation}

Moments of the variable $m_n$ are obtained from the generating function
\cite{gf} satisfied by the Stirling numbers of the first kind \cite{gkp}
\begin{equation}
\label{stirling}
S_n(w)=\sum_{m=0}^n {n\brack m} w^m=w(w+1)\ldots(w+n-1).
\end{equation}
Taking the logarithmic derivative gives the average
\begin{equation}
\label{m-av}
\langle m_n\rangle=\frac{d}{dw}\,\ln S_n(w)|_{w=1} = H_{n}
\end{equation}
in terms of the harmonic numbers $H_n=\sum_{j=1}^n\frac{1}{j}$.  Using the
large $n$ asymptotics of the harmonic numbers \cite{gkp}, we conclude that
the average grows logarithmically
\begin{equation}
\label{m-av-large}
\langle m_n\rangle=\ln n+\gamma+\frac{1}{2n}+\cdots.
\end{equation}

The second derivative \hbox{$\frac{d^2}{dw^2}\ln S_n(w)|_{w=1}$} similarly
gives $\langle m_n(m_n-1)\rangle$. The variance, $w_n^2=\langle m_n^2\rangle
-\langle m_n\rangle^2$, follows
\begin{equation}
\label{var}
w_n^2=H_n-H_n^{(2)}.
\end{equation}
Here, $H_n^{(2)}=\sum_{j=1}^n\frac{1}{j^2}$ are the second-order
harmonic numbers.  Asymptotically, the variance grows logarithmically
\begin{equation}
\label{var-large} w_n^2=\ln
n+\gamma-\frac{\pi^2}{6}+\frac{3}{2n}+\cdots.
\end{equation}

The leading asymptotic behavior of the distribution can be evaluated
as well.  Using properties of the Stirling numbers, the distribution
for small $m$ reads
\begin{eqnarray}
\label{small-m}
P_{n,1}&=&\frac{1}{n},\nonumber\\
P_{n,2}&=&\frac{1}{n}\,H_{n-1},\\
P_{n,3}&=&\frac{1}{2n}\,\left[H_{n-1}^2-H_{n-1}^{(2)}\right].\nonumber
\end{eqnarray}
These exact results reflect that the distribution is Poissonian for
sufficiently small $m$:
\begin{equation}
\label{poisson}
P_{n,m}\simeq \frac{1}{n}\,\frac{(\ln n)^{m-1}}{(m-1)!}.
\end{equation}

The Poissonian form corresponds to the small-$m$ tail of the distribution. To
obtain the distribution for typical, rather than extremal, values of $m$, we
consider the continuum limit of the recursion relation (\ref{dis-rec}) where
the distribution satisfies
\begin{equation}
\label{cont}
n\frac{\partial P}{\partial n}+\frac{\partial P}{\partial m}
=\frac{1}{2}\,\frac{\partial^2 P}{\partial m^2}.
\end{equation}
The change of variables $n\to t=\int_1^n dn'/n'$ transforms (\ref{cont}) into
the standard diffusion-convection equation whose solution admits
a Gaussian form 
\begin{equation}
\label{gaussian}
P_{n,m}\to \frac{1}{\sqrt{2\pi w_n^2}}\,
\exp\left[-\frac{(m_n-\langle m_n\rangle)^2}{2w_n^2}\right].
\end{equation}
As $m_n=\ln x_n/\ln 2$, the distribution of $x_n$ is therefore log-normal.
Moreover, the variance of the random variable $\ln x_n$ is simply $(\ln
2)^2\ln n$. 

Both the Poissonian and the Gaussian behaviors follow from the more general
asymptotic form of $P_{n,m}$ 
\begin{equation}
\label{modified}
P_{n,m}\simeq 
\frac{1}{\Gamma(m/\ln n)}\,
\frac{(\ln n)^m}{n\cdot m!}
\end{equation}
that holds when $m\to\infty$ and $n\to\infty$ with the ratio $m/\ln n$
being finite. Indeed, using the asymptotic relation $\Gamma(x)\to
x^{-1}$ as $x\to 0$ one recovers (\ref{poisson}); the peak of the
distribution (\ref{modified}) is at $m=\ln n$ and expansion in the
vicinity of this peak recovers (\ref{gaussian}). We term the
distribution (\ref{modified}) the modified Poissonian distribution.

To derive (\ref{modified}), we use (\ref{dis-sol})--(\ref{stirling}) and the
Cauchy theorem to express $P_{n,m}$ as an integral
\begin{equation}
\label{contour}
P_{n,m}= \frac{1}{2\pi i}\oint
\frac{dw}{w^{m+1}}\,\frac{w(w+1)\ldots(w+n-1)}{n!}
\end{equation}
over an arbitrary simple closed contour enclosing the origin in the
complex $w$ plane. When $n\to\infty$, the contour integral is easily
computed by applying the steepest descent method. The saddle point is
determined from
\begin{equation}
\label{saddle}
\frac{m}{w_*}=\sum_{j=1}^n\frac{1}{w_*+j}\,.
\end{equation}
Asymptotically, $w_*\simeq m/\ln n$. 
We now deform the integration contour to the contour of steepest descent that
runs through the saddle point along the imaginary axis (in the complex $w$
plane). Writing $w=w_*(1+iy)$ and taking into account that the dominant
contribution is gathered near $y=0$, we obtain
\begin{eqnarray*}
P_{n,m} &\simeq& \frac{w_*}{2\pi}\,
\frac{(w_*+1)\ldots(w_*+n-1)}{w_*^m\cdot n!}\int_{-\infty}^\infty
dy\,e^{-m y^2/2} \\
&=&\frac{1}{\sqrt{2\pi m}}\, \frac{\Gamma(w_*+l)}{w_*^{m}\,\Gamma(n+1)\,
\Gamma(w_*)}\\
&\simeq&\frac{1}{\sqrt{2\pi m}}\, \frac{n^{w_*-1}}{w_*^{m}\,\,\Gamma(w_*)}
\end{eqnarray*}
where we used two properties of the gamma function --- the difference
equation $\Gamma(x+1)=x\,\Gamma(x)$ and the asymptotic relation
$\frac{\Gamma(x+a)}{\Gamma(x)}\to x^a$ as $x\to\infty$.  Inserting
$w_*\simeq m/\ln n$ and using the Stirling formula leads to
(\ref{modified}).

\subsection{The Typical Behavior}

The asymptotic distribution (\ref{gaussian}) and the growth law
(\ref{m-av-large}) lead to the typical behavior
\begin{equation}
\label{typical}
x_n\simeq C\,n^{\ln 2}
\end{equation}
with $C=2^\gamma\cong 1.491967$ and $\gamma\cong 0.577215$ the Euler's
constant.  Since asymptotically, $\ln x_n\to \langle \ln x_n\rangle$,
the typical behavior (\ref{typical}) emerges from the $s\to 0$ limit
of the properly modified moments $\langle x_n^s\rangle^{1/s}$ in
Eq.~(\ref{moments}). In other words, the Lyapunov exponent
$\lambda=\ln 2$, defined via $x_n\sim \exp(\lambda \ln n)$ is obtained
from the moment spectrum using $\lambda=\lim_{s\to 0}s^{-1}\beta(s)$.
However, the typical behavior (\ref{typical}) and the asymptotic
distribution (\ref{gaussian}) do not yield the moments as they imply
the quadratic moment spectrum $s\ln 2+ \frac{1}{2}(s\ln 2)^2$, equal
to the first two terms in the Taylor expansion of $\beta(s)$. 

There are large fluctuations between successive elements in a given
series and large series-to-series variations. The typical behavior is
eventually approached but very slowly, as illustrated in Fig.~1.

\begin{figure}[t]
\centerline{\epsfxsize=8cm\epsfbox{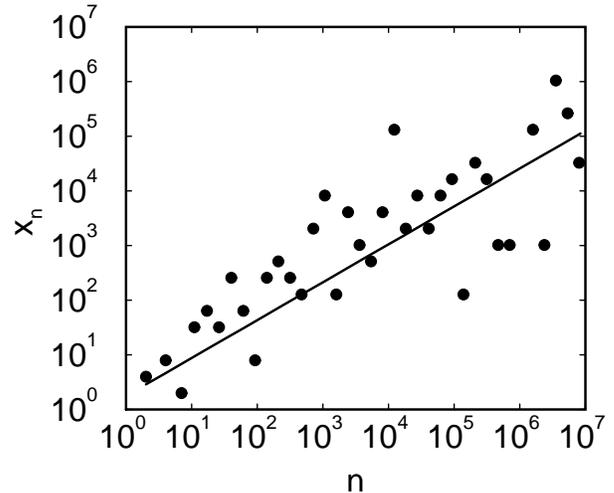}} 
\caption{A single realization of the random geometric series (bullets)
versus the typical behavior (\ref{typical}), shown using a line. For
clarity, only a small fraction of the series elements are displayed.}
\end{figure}

Let us compare random geometric series with random Fibonacci-like series.
Overall, the behavior is in line with the behavior found for the
series $x_n=x_{n-1}+x_p$ with $0\leq p\leq n-1$. In both cases, the
distribution of $x_n$ is log-normal and the moments exhibit
multiscaling \cite{bk}. In the present case, it is possible to find
the Lyapunov exponent. However, the behavior is unlike the one found
for the random sequence $x_n=x_p+x_q$ with \hbox{$0\leq p,q\leq n-1$}
despite the fact that in both cases the average is $\langle x_n\rangle
=n+1$.  The average characterizes all the moments and the distribution
approaches an ordinary scaling from $P_n(x)\to n^{-1}\Phi(xn^{-1})$
\cite{bk}.

\subsection{Extremal Statistics}

The span of the additive random sequence, i.e. the set of all possible
values of $m$, provides an additional statistical characterization. In
every realization, this set contains no gaps, so the span is
equivalent to the largest sequence element $M_n$. Thus, the span is
directly related to extremal characteristics of the sequence. To find
how $M_n$ grow with $n$, it is necessary to consider the large $m$
tail of the probability distribution outside the Gaussian region. The
modified Poissonian distribution (\ref{modified}) suggests that
$M_n\sim \ln n$. 

We obtain the growth of the maximal element in the series heuristically
using the extreme value criterion
\begin{equation}
\label{span}
\sum_{n'=M_n}^n \sum_{m'=M_n}^{n'} P_{n',m'}\sim 1. 
\end{equation}
Since the the distribution $P_{n,m}$ quickly diminishes with $m$ in the 
tail region, this extreme statistics criterion becomes 
$\sum_{n'=M_n}^n P_{n',M_n}\sim 1$, and  as $M_n\ll n$, one has
\begin{equation}
\label{span-1}
nP_{n,M_n}\sim 1.
\end{equation}
Combining this criterion with the distribution (\ref{modified}) and
using the Stirling formula, we arrive to the following asymptotic
growth of the maximal value
\begin{equation}
\label{mn}
M_n\simeq e\ln n.
\end{equation}
Numerical simulations are in good agreement with this estimate
(Fig.~2).  A more rigorous derivation including the leading correction
is given in Appendix A.

\begin{figure}[t]
\centerline{\epsfxsize=8cm\epsfbox{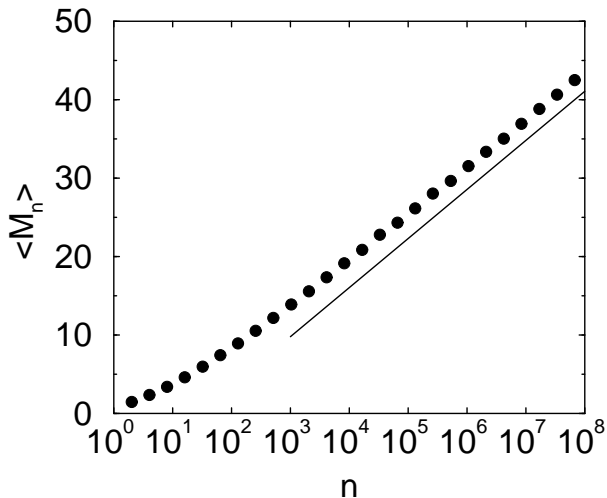}} 
\caption{The maximal element in the series. The average $\langle
M_n\rangle$, obtained from $100$ independent realizations of the
random additive series (bullets) is compared with the heuristic
estimate (\ref{mn}), shown using a line of slope $e$.}
\end{figure}

This growth is ultimately connected with the frequency by which the
largest element in the sequence occurs. At the $n$th step, the maximal
value $M_n$ is augmented by one with probability $h/n$ with $h$ the
frequency by which the largest element occurs. Thus, $\langle
M_n\rangle=\langle M_{n-1}\rangle+\frac{\langle h\rangle}{n}$. This
leads to the growth law $\langle M_n\rangle\simeq \langle h\rangle \ln
n\,$. The frequency $h$ is a random variable that in principle depends on
$n$ yet it has a stationary distribution $p(h)$ in the large $n$
limit. The growth law (\ref{mn}) implies $\langle h\rangle =\sum h p(h)=e$.

\section{Generalizations}

There are a number of natural generalizations of the stochastic
sequences (\ref{rule-mult}) and (\ref{rule-add}). Below, we briefly
describe two examples. In both cases we consider the variable $m_n$
directly.

\subsection{Random Random Walk}

In the random geometric series, the previous element is chosen
randomly while the recursive rule is deterministic. We thus consider
the {\em stochastic} recursion relation
\begin{equation}
\label{rule-rw}
m_n=m_p\pm 1
\end{equation}
with $m_0=0$ and a randomly chosen $0\leq p\leq n-1$ on the $n^{\rm th}$
step. We assume that both signs in (\ref{rule-rw}) are taken with equal
probability.  We term the sequence generated by (\ref{rule-rw}) the random
random walk.

The moments can be obtained recursively, as in the random geometric
series and we merely quote the results. The first moment vanishes,
$\langle m_n\rangle=0$, and the variance is given by the harmonic numbers
\begin{equation}
\label{m2-rw}
w_n^2=H_n.
\end{equation}
The asymptotic behavior is therefore $w_n^2\simeq \ln n$.

The probability $P_{n,m}$ that the walk is at position $m$ at the
$n^{\rm th}$ step obeys
\begin{equation}
\label{dis-eq-rrw}
P_{n,m}=\frac{n-1}{2n}P_{n-1,m}
+\frac{1}{2n}\left[P_{n-1,m-1}+P_{n-1,m-1}\right].
\end{equation}
Taking the continuum limit, we find that the distribution satisfies
$n\,\frac{\partial P}{\partial n}=\frac{\partial^2 P}{\partial m^2}$. This
diffusion equation shows that the distribution is Gaussian
\begin{equation}
\label{gaussian-rw}
P_{n,m}\simeq {1\over \sqrt{2\pi w_n^2}}\exp\left[-{m^2\over 2w_n^2}\right].
\end{equation}
Therefore, all moments are characterized by the variance: $\langle
m_n^{2k}\rangle \simeq (2k-1)!!(\ln n)^k$.  The random random walk
spreads very slowly with the typical spread
\begin{equation}
\label{typical-rw}
m\sim \sqrt{\ln n}.
\end{equation}
Hence, the first passage time, the time to reach a site of distance
$m$ from the origin grows as $\exp(m^2)$. What remains an open
question is whether the distribution $P_{n,m}$ can be obtained in a
closed form from the generating function $\sum_{n,m}z^n w^m
P_{n,m}=(1-z)^{-w(w+1)}$. In terms of the variable $x_n=2^{m_n}$, the
growth is slower than any power law, $x_n\sim \exp(\ln 2\sqrt{\ln n})$.

\subsection{Two Dimensions}

Thus far, we considered only one-dimensional sequences. In physical systems,
it is generally believed that disorder, no matter how small, is always
relevant in two-dimensions \cite{cpv,jml}. However, disordered
two-dimensional systems are typically untreatable analytically.

We consider a natural generalization of (\ref{rule-add}) to a
two-dimensional square lattice.  Starting from $m_{\bf 0}=1$,
where ${\bf 0}=(0,0)$ is the site at the origin, the values at
further sites are determined recursively according to
\begin{equation}
\label{rule-2d}
m_{\bf n}=m_{\bf p}+1
\end{equation}
Here ${\bf p}=(p_1,p_2)$ is chosen equiprobably among lattice sites
that are closer to the origin than ${\bf n}$, i.e. $|{\bf p}|<|{\bf
n}|$;. We choose the ``manhattan distance'' from the origin $|{\bf
n}|=|n_1|+|n_2|$ as the measure of distance.

The probability distribution depends only on the norm $n=|{\bf n}|$,
so we keep the notation $P_{n,m}$. For the norm $n=|{\bf
n}|=|n_1|+|n_2|$, there are $4n$ lattice sites a distance $n$ from the
origin, $1+4+\ldots +4(n-1)=1+2n(n-1)$ lattice sites which are a 
distance $\leq n-1$ from the origin.  The probability distribution
satisfies the recursion relation
\begin{eqnarray}
\label{dis-eq-2d}
P_{n,m}&=&\left[1-\frac{4(n-1)}{1+2n(n-1)}\right]P_{n-1,m}\\
&+&\frac{4(n-1)}{1+2n(n-1)}\,P_{n-1,m-1}\nonumber
\end{eqnarray}
for $n\geq 2$ with $P_{0,m}=\delta_{m,0}$ and $P_{1,m}=\delta_{m,1}$.

In analogy with the one-dimensional case, we write the distribution in
the form
\begin{equation}
\label{Pnm2-sol}
P_{n,m}=\frac{1}{\Pi_n}\,G_{n,m}^{(2)}
\end{equation}
with $\Pi_n=\prod_{j=1}^n [1+2j(j-1)]$ and $G_{n,m}^{(2)}$ the
two-dimensional analogs of the Stirling numbers of the first
kind. These non-negative integer numbers obey the fundamental
recursion relation
\begin{eqnarray}
G_{n+1,m}^{(2)}=[1+2n(n-1)]G_{n,m}^{(2)}+4nG_{n,m-1}^{(2)}
\end{eqnarray}
for $n\geq 2$ and $G_{0,m}^{(2)}=\delta_{m,0}$, $G_{1,m}^{(2)}=\delta_{m,1}$.
The corresponding generating function is
\begin{equation}
\label{stirling2}
\sum_{m=0}^n G_{n,m}^{(2)}\,w^m=w\prod_{j=1}^n [1+2(j-1)(j-2+2w)]
\end{equation}
for $n\geq 1$. Using this generating function, the average and the
variance are
\begin{eqnarray}
\langle m_n\rangle&=&1+\sum_{j=1}^n \frac{4(j-1)}{1+2j(j-1)}\,,\\
w_n^2&=&\sum_{j=1}^n \frac{4(j-1)}{1+2j(j-1)}
-\sum_{j=1}^n \left[\frac{4(j-1)}{1+2j(j-1)}\right]^2.\nonumber
\end{eqnarray}
The leading asymptotic behaviors are $\langle m_n\rangle\simeq 2\ln n$
and $w_n^2\simeq 2\ln n$.  In the continuum limit, the distribution
obeys the diffusion-convection equation that now has the form
$n\,\frac{\partial P}{\partial n}+2\frac{\partial P}{\partial m}
=\frac{\partial^2 P}{\partial m^2}$. Asymptotically, the distribution
is Gaussian as in the one-dimensional case (\ref{gaussian}) with the
average and the variance merely modified by the factor $2$. One can
also show, by generalizing the moment recursion relation
(\ref{mom-rec}), that the spectrum is also modified by the same
factor: $\beta(s)=2(2^s-1)$.

\section{Summary}

In summary, we considered random sequences where an element depends on
a previous randomly chosen element and have shown that they exhibit a
similar phenomenology as sequences that involve dependence on a few
previous elements. The typical behavior and the moment behavior
provide a statistical characterization of the sequence.  The growth
laws depend sensitively on details of the recurrence relations.

We obtained a number of exact and asymptotically exact results for the
probability distribution and its moments. For the random geometric
series, the sequence growth is algebraic. The moments exhibit
multi-scaling asymptotic behavior and also contain information
regarding the typical behavior. Asymptotically, the probability
distribution becomes log-normal but it does not fully characterize the
actual moment behavior.

There are additional interesting questions that can be asked for this
family of random series including the likelihood of monotonically
increasing sequences, growth of correlations between two different
elements in the same sequence, and statistics of the number of
distinct elements in a given sequence.

\acknowledgments We thank Lan Yuehen for useful comments. This
research was supported in part by DOE(W-7405-ENG-36).

\appendix
\section{Derivation of Eq.~(\ref{mn})}

Substituting (\ref{modified}) into the criterion \hbox{$\sum_{n'=M_n}^n
P_{n',M_n}\sim 1$} and replacing the summation by integration yields 
\begin{eqnarray}
\label{Pnm-sum}
\sum_{n'=M_n}^n P_{n',M_n} &\simeq& \frac{1}{(M_n)!}\int^n \frac{dn'}{n'}\,
\frac{(\ln n')^{M_n}}{\Gamma(M_n/\ln n')}\nonumber\\
&\sim& \frac{(\ln n)^{M_n+1}}{(M_n+1)!}\sim 1. 
\end{eqnarray}
Taking the logarithm of (\ref{Pnm-sum})  and using the Stirling
formula, we obtain an implicit relation for the maximal element
\begin{equation}
\label{mu}
M_n-M_n \ln\left(\frac{M_n}{\ln n}\right)=\frac{3}{2}\,\ln M_n - \ln\ln n.
\end{equation}
This yields the leading correction to Eq.~(\ref{mn})
\begin{equation}
\label{mn-as}
M_n\to e\ln n -\frac{1}{2}\,\ln \ln n+\ldots.
\end{equation}
$M_n$ is of course a random variable and (\ref{mn-as}) represents its
average $\langle M_n\rangle$. We anticipate that fluctuations in $M_n$
remain finite as $n\to\infty$.


\begin{thebibliography}{99}
\bibitem{gf}
     Alternatively, it can be derived from the double generating function
     $\sum_{n,m}z^nw^mP_{n,m}=(1-z)^{-w}$.
\end{thebibliography}
\end{document}